\documentclass{emulateapj}
\slugcomment{Accepted for publication in ApJ Letters }

\shorttitle{TRACING THE GALACTIC THICK DISK TO SOLAR METALLICITIES}
\shortauthors{BENSBY ET AL.}

\newcommand\teff{T_{\rm eff}}

\newcommand\ulsr{U_{\rm LSR}}
\newcommand\vlsr{V_{\rm LSR}}
\newcommand\wlsr{W_{\rm LSR}}

\begin{document}

\title{
Tracing the Galactic thick disk to solar metallicities\altaffilmark{1}
}
\author{
T. Bensby,\altaffilmark{2}
A.R. Zenn,\altaffilmark{2}
M.S. Oey,\altaffilmark{2}
and
S. Feltzing\,\altaffilmark{3}
}

\altaffiltext{1}{Based on data collected with the 6.5\,m Magellan
telescopes at the Las Campanas Observatory, and with the Very
Large Telescope at the European Southern Observatory
(ESO proposal 72.B-0179).}
\altaffiltext{2}{Department of Astronomy, University of Michigan, 
Ann Arbor, MI, USA; {\tt tbensby@umich.edu}}
\altaffiltext{3}{Lund Observatory, Lund, Sweden;
{\tt sofia@astro.lu.se}}

\begin{abstract}
We show that the Galactic thick disk reaches at least solar 
metallicities, and that it experienced strong chemical enrichment
during a period of $\sim3$\,Gyr, ending around 8-9\,Gyr ago. 
This finding puts further constraints on 
the relation and interface between the thin and thick disks, and their 
formation processes.
Our results are based on a detailed elemental abundance analysis of 261 
kinematically selected F and G dwarf stars in the solar neighborhood:
194 likely members of the thick disk and 67 likely members of the thin disk, 
in the range $\rm -1.3\lesssim[Fe/H]\lesssim+0.4$. 
\end{abstract}

\keywords{
Galaxy: disk ---
Galaxy: formation ---
Galaxy: evolution ---
solar neighbourhood ---
stars: abundances ---
stars: kinematics
}

\section{Introduction}

The Milky Way has since the early 1980s been known to have two disk 
components, a thin disk and a thick disk \citep{gilmore1983}.
Since then, several studies, using high-resolution spectra to derive
elemental abundances in disk dwarf stars, have been aimed at establishing
the properties of the thick disk, and to better understand
its origin and role in our Galaxy
\citep[e.g.,][]{gratton2000,prochaska2000,fuhrmann2004,mishenina2004,
reddy2006,bensby2003,bensby2005}.
The thick disk is now known to be a major Galactic stellar 
population, and that its stars have hotter kinematics, higher ages, 
and are chemically distinct from the stars of the thin disk.
All this points to separate origins and different chemical histories
for the thin and thick  disks.

Recently, a lot of structure has been observed amongst the stars in the Galaxy.
In the disk in the solar neighbourhood this is seen as various stellar streams and 
moving groups \citep[e.g.][]{famaey2005short,helmi2006short}; 
and at larger distances, features such as e.g. 
"The Field of streams" \citep[e.g.][]{belokurov2006short}
have been detected. So, did the thick disk form as a single 
entity in the initial collapse of the protogalactic cloud 
\citep[e.g.,][]{eggen1962}, and/or is it a result of an ancient merger event, 
or is it made up of a stars coming from
streams and merger debris, i.e. a hierarchical origin 
\citep[e.g.][]{abadi2003,brook2004,robertson2004}?
A persistent question is why the Milky Way has two
disk populations.

Abundance trends and the metallicity distribution function 
of the thick disk are vital records to its formation and evolution.
However, the high metallicity limit of the thick disk remains poorly defined. 
For instance, \cite{fuhrmann2004,mishenina2004,reddy2006} 
suggest that the thick disk extends only up to $\rm [Fe/H]\,\approx\,-0.3$,
because their candidate thick disk stars at higher [Fe/H] either fall
within their thin disk abundance trends and/or have
highly eccentric orbits that are near the Galactic plane.
Hence their origins should be sought elsewhere, perhaps in stellar streams like 
the Hercules stream \citep[see e.g.][]{famaey2005short}. But, even if possible 
Hercules stream stars are weeded out, stars with thick-disk-like kinematics 
at high [Fe/H] still remain 
\citep[][see also Fig.~\ref{fig:contour}]{soubiran2005,bensby2007letter}. 
Furthermore, in \cite{bensby2003,bensby2004,bensby2005,bensby2006}
we find that the thick disk stars differ significantly from the thin disk 
stars, both in terms of abundance ratios as well as stellar ages, 
even at [Fe/H] close to solar. However, those results are based on a small 
stellar sample and need confirmation.

As described, the current data for the metal-rich thick disk are
confusing and ambiguous. It is therefore necessary to isolate the thick disk
abundance relations from those of other populations.
Therefore, we have carried out an extensive spectroscopic survey of 
metal-rich stars that are kinematically associated with the Galactic thick 
disk. In this Letter, we discuss Ti and Ba abundance trends, and combine 
our new results with our thin and thick disk results from
\cite{bensby2003,bensby2005}. Other $\alpha$-, $r$-, $s$-, and iron peak 
elements will be discussed in an upcoming paper (Bensby et al., in prep) 
together with the details of the kinematic selection criteria and the 
abundance analysis.

\section{Selection of targets, observations and abundance analysis}
\label{sec:observations}

\begin{figure}
\resizebox{\hsize}{!}{
\includegraphics[bb=18 215 592 650,clip]{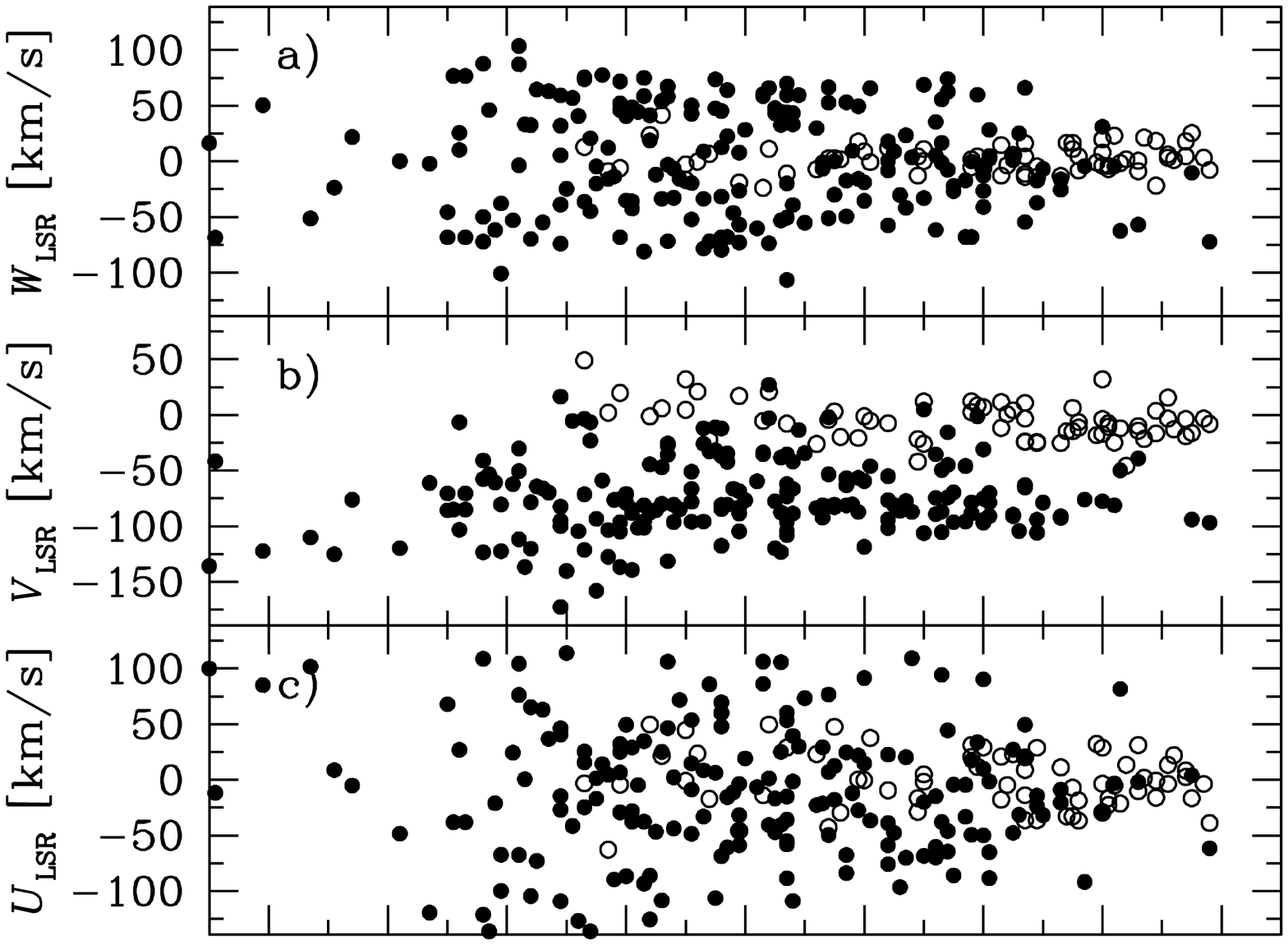}}
\resizebox{\hsize}{!}{
\includegraphics[bb=18 170 592 465,clip]{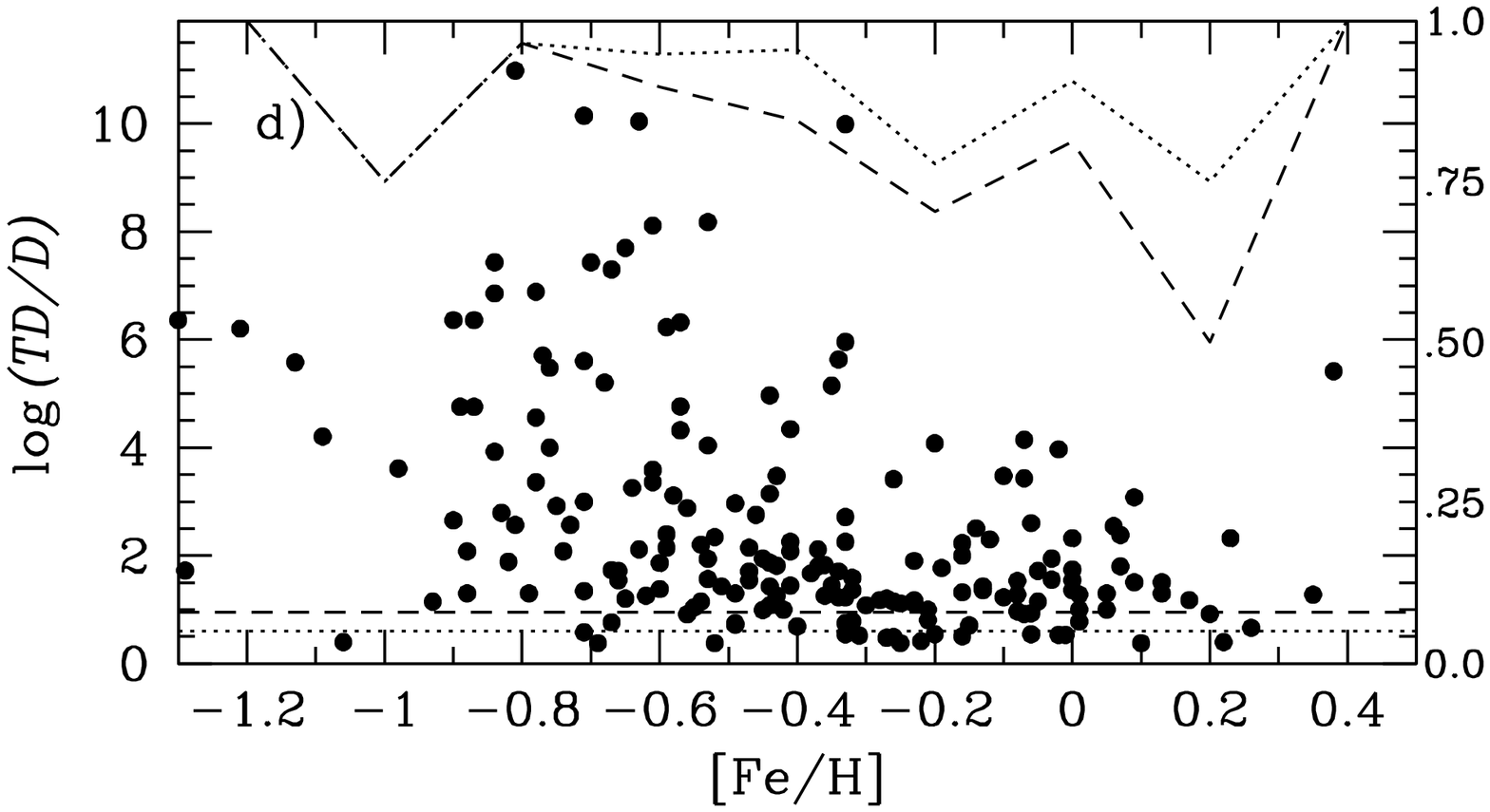}}
\caption{
        {\sl (a)-(c):} Velocity-metallicity plots
        for the stellar sample. Thin disk and
        candidate thick disk stars are marked by
        open and filled circles, respectively.
	{\sl (d):} $TD/D$ probability ratios versus metallicity
	for the thick disk candidates. Bottom lines indicate
	$TD/D=4$ (dotted line) and $TD/D=9$ (dashed line).
	Lines on top give the corresponding fraction of stars
	that have $TD/D$ larger than these ratios (scale on the
	right hand side).
        [Fe/H] values are from our spectroscopic work.
\label{fig:contour}}
\end{figure}

The kinematic method from \cite{bensby2003,bensby2005} was used to select 
possible thick disk F and G dwarf stars from the \cite{nordstrom2004short} 
catalogue. Briefly, the method assumes Gaussian velocity distributions
for all stellar populations, and that the solar neighbourhood can be
represented as a mixture of only the thin disk, the thick disk, the Hercules 
stream, and the halo. Candidate thick disk stars are selected as 
those that have probabilities of belonging to the thick disk that are at least 
twice the probabilities of belonging to any of the other populations 
(and likewise for the other populations). The space velocities for the 159 
new thick disk and 10 new thin disk stars are shown in 
Fig.~\ref{fig:contour}a-c together with 35 thick 
disk and 57 thin disk stars from \cite{bensby2003,bensby2005}.
Also shown, in Fig.~\ref{fig:contour}d-e, is how the thick disk-to-thin disk
probability ratios ($TD/D$) vary with [Fe/H]. 

Echelle spectra were obtained in 2005 and 2006 with the MIKE spectrograph,
on the Magellan Clay 6.5\,m telescope, for 145 new thick disk stars
($R\approx65\,000$, $S/N\gtrsim250$),
and in 2004 with the UVES spectrograph, on the ESO Very Large Telescope,
for 14 new thick disk and 10 new thin disk stars
($R\approx110\,000$, $S/N\gtrsim250$).

The MARCS model stellar atmospheres 
\citep{gustafsson1975,edvardsson1993short,asplund1997} were used in the abundance
analysis. Excitation balance, and balance with line strength, of 
abundances from Fe\,{\sc i} lines, were used to determine effective 
temperatures and the microturbulence parameter. 
For the surface gravities we exploited accurate distances based 
on {\sc\sl Hipparcos} parallaxes \citep{esa1997}. Final abundances were
normalised on a line-by-line basis with our solar values as reference and
then averaged for each element.

Stellar ages were determined with the help of the Yonsei-Yale (Y$^{2}$)
isochrones \citep{kim2002,demarque2004}, with appropriate $\alpha$-enhancements,
in the $\teff$-$M_{\rm V}$ plane.
Upper and lower limits on the ages were estimated from the error bars due
to an uncertainty of $\pm 70$\,K in $\teff$
and the uncertainty in $M_{\rm V}$ due to the error in the parallax
\citep[see also][]{bensby2003}.

\section{Results and discussion} \label{sec:results}

\subsection{Abundance trends}

\begin{figure*}
\resizebox{\hsize}{!}{
\includegraphics[bb=18 165 592 655,clip]{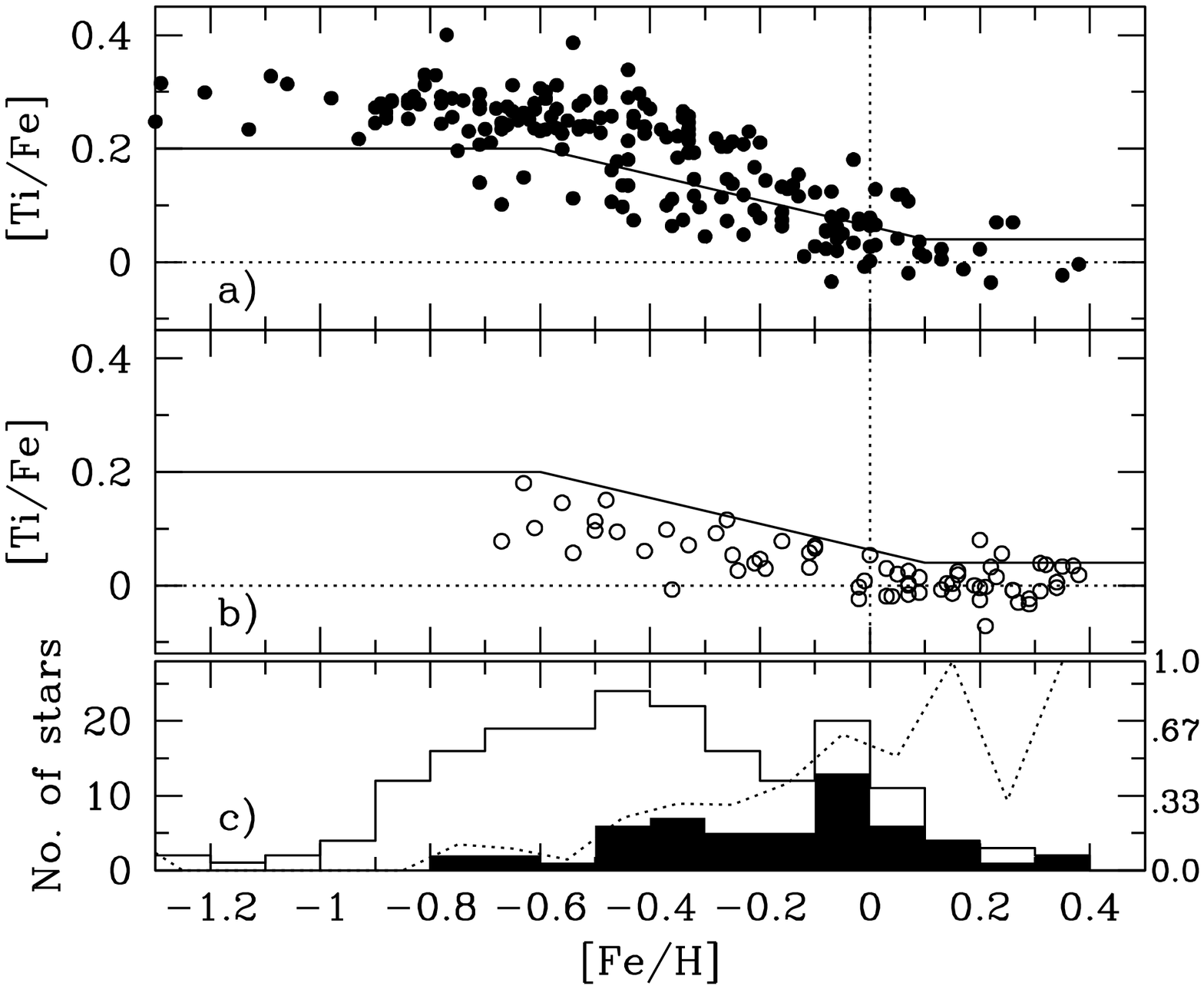}
\includegraphics[bb=18 165 592 655,clip]{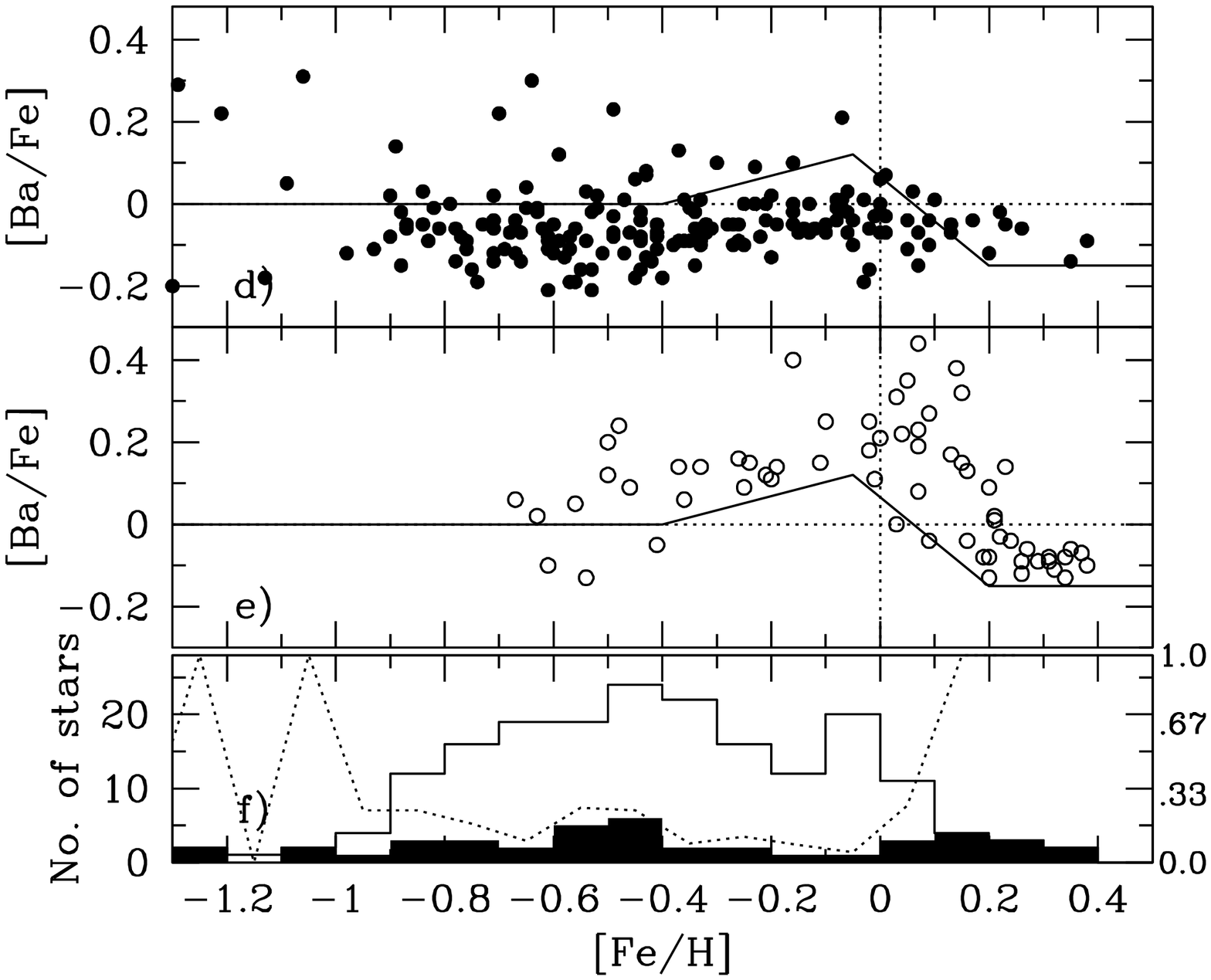}}
\caption{
        {\sl Top four panels:}
        [Ti/Fe] and [Ba/Fe] versus [Fe/H]. Thin and thick disk stars are
        marked by open and filled circles, respectively.
        Solid lines marks the boundary for the thin sample.
        {\sl Bottom panels:}
        Distribution of all 195 thick disk stars
        {\sl (white histograms)}, and the thick disk stars that
        fall within the thin disk abundance trends {\sl (black histograms)}.
        The dotted line in the bottom panels shows the fraction (scale on the
        right-hand side) of thick disk stars in each
        bin that fall within the thin disk abundance trends.
\label{fig:haltplottar0}}
\end{figure*}

Figure~\ref{fig:haltplottar0} shows the resulting [Ti/Fe] and [Ba/Fe] 
versus [Fe/H] trends. The stars associated with the 
thick disk first show a [Ti/Fe] plateau at $\rm [Fe/H] \lesssim -0.4$, 
a signature of fast enrichment from massive stars. At higher metallicities, 
the thick disk [Ti/Fe] ratio declines, indicating the delayed enrichment 
from SN\,Ia. The thin disk shows an overall 
shallow decline  in [Ti/Fe], characteristic of 
slow enrichment by {\sl both} massive and low-mass stars. 
At $\rm [Fe/H]\approx0$, the trends for the two disks converge.

[Ba/Fe] for the thick disk evolves almost in lockstep with [Fe/H].
As solar metallicity is approached, the thin and thick disk [Ba/Fe] 
trends diverge. At $\rm [Fe/H] >0$ it again becomes 
hard to differentiate the two disks.

Both [Ti/Fe] and [Ba/Fe] versus [Fe/H] 
demonstrate that kinematically hot stars associated with the thick disk
extend to solar metallicities. However, it is also evident that
there are thick disk stars 
that do not follow the general thick disk abundance trends. Instead, they
chemically behave as thin disk stars. This is at least
evident in the [Ti/Fe]-[Fe/H] plot as $\rm [Fe/H]\approx0$ is approached.

To try to determine the nature of these ambiguous stars, we use 
the thin disk sample (Figs.~\ref{fig:haltplottar0}b and e) to visually 
define boundaries on [Ti/Fe] and [Ba/Fe] for the thin disk 
(shown as solid lines in the {\sl upper four panels} of 
Fig.~\ref{fig:haltplottar0}). 
The number of candidate thick disk stars that fall within the thin disk 
abundance trends are shown in the {\sl bottom two panels} of 
Fig.~\ref{fig:haltplottar0}. There is a steady increase with metallicity of candidate 
thick disk stars that fall within the thin disk [Ti/Fe] trend, suggesting that 
the contamination from the high-velocity tail of the thin disk increases with 
[Fe/H]. The fraction that fall within the thin 
disk [Ba/Fe]-[Fe/H] trend is, on the other hand, generally small, and 
with no apparent trend.  This suggests that essentially all candidate thick 
disk stars could be genuine thick disk stars.

Due to the closeness of the thin and thick disk [Ti/Fe] trends at higher [Fe/H]
one can expect true members of the thick disk to fall within the thin disk 
trend, and vice versa. And, since the Ba abundances are based on only 3-4 
Ba\,{\sc ii} lines, there are larger measurement uncertainties in [Ba/Fe] 
than in [Ti/Fe]. Ba abundances could also be influenced by 
NLTE effects, hyperfine and isotopic structure, and blends from other 
spectral lines \citep[see, e.g.,][]{mashonkina2006}; effects that we have not 
accounted for. On the other hand, we present a strictly
differential abundance analysis. If the above effects were severe, 
we would not find well-defined and distinct Ba trends for 
two kinematically selected samples. Hence, we believe that our Ba abundances
are well determined. 

\subsection{Age trends}

\begin{figure}
\resizebox{\hsize}{!}{
\includegraphics[bb=18 165 592 670,clip]{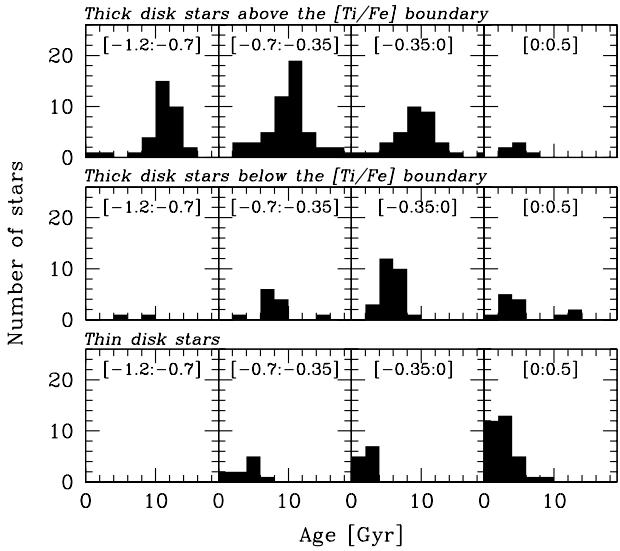}}
\caption{
        Top and middle panel show age distributions for
        thick disk stars that fall above and below the [Ti/Fe] boundary,
        respectively. Bottom panel shows the thin disk distribution.
	All age distributions are divided into four metallicity bins,
	as indicated in the square brackets at the
	top of each panel.
\label{fig:agemess}}
\end{figure}

\begin{figure}
\resizebox{\hsize}{!}{\includegraphics[bb=18 165 592 718,clip]{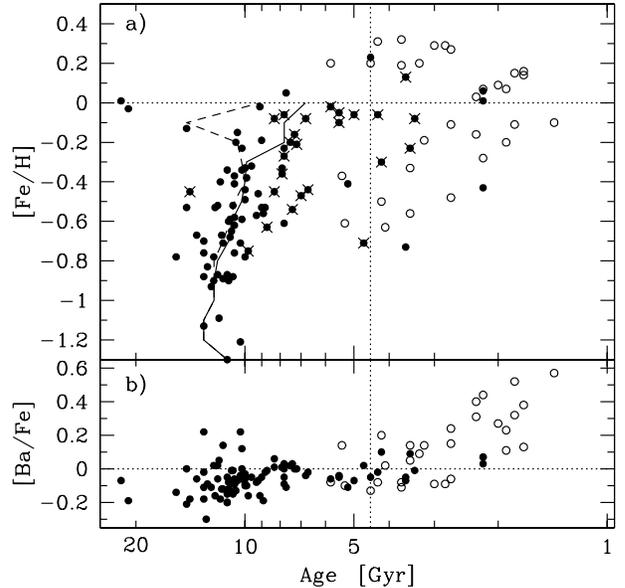}}
\caption{
        [Fe/H] and [Ba/Fe] versus age. Stars for which the upper and
        lower age estimates differ by at most 35\,\% are included.
        Thin and thick disk stars are marked by open and filled circles,
        respectively. The solid line  shows the running
        median ages (see text) for the thick disk stars, and the dashed line 
        when excluding thick disk stars that have thin disk [Ti/Fe]
        values (marked by crosses).
        The vertical, dotted line, is the age of the Sun (4.5\,Gyr).
\label{fig:agefeh}}
\end{figure}

The top panel in Fig.~\ref{fig:agemess} shows the age distributions for the 
candidate thick disk stars that follow the thick disk [Ti/Fe] trend
(as defined by the thin disk boundary line in Fig.~\ref{fig:haltplottar0}a)
while the middle panel shows those that do not. Each sub-sample has been 
divided into four metallicity bins, as shown. The bottom panel shows the age 
distributions for the thin disk sample. 

The candidate thick disk stars with thin disk [Ti/Fe] ratios appear to be 
younger than those above the boundary. 
For instance, in the $\rm -0.35<[Fe/H]<0$ bin, only one out of 26 candidate 
thick disk stars (4\,\%) that have a thin disk [Ti/Fe] ratio is older than 
8\,Gyr. In the same metallicity bin, 23 out of 33 stars (70\,\%)
that remain above the thin disk [Ti/Fe] boundary are older than 8\,Gyr.
This duality in both ages and abundances again points to two distinct 
Galactic disk populations, both reaching $\rm [Fe/H]=0$.

Figure~\ref{fig:agefeh} shows [Fe/H] and [Ba/Fe] as a
function of age, excluding stars with estimated
upper and lower age limits (see Sect.~\ref{sec:observations}) that differ by 
more than 35\,\%. Running medians of the ages for the thick disk, 
calculated in steps of 0.1\,dex in [Fe/H], using a 0.2\,dex wide window in
[Fe/H], both with and without thick disk stars that have thin 
disk [Ti/Fe] ratios, are shown in Fig.~\ref{fig:agefeh}a.
For $\rm [Fe/H]\lesssim-0.8$, median ages are typically $\sim 12$\,Gyr.
The median age at higher [Fe/H] depends on whether
thick disk candidates that have thin disk [Ti/Fe] ratios
are included or not.  As many of the stars that fall below the 
[Ti/Fe] boundary have ages comparable to the stars that do not, 
it is likely that the actual relation
is intermediate to the solid and dashed lines. The age of the 
thick disk at solar metallicities is then $\sim9$\,Gyr, i.e. it takes the 
thick disk $\sim3$\,Gyr to reach $\rm [Fe/H]\approx0$.

\subsection{The metal-rich end of the Galactic thick disk}

Our kinematically hot stars appear to come from an old stellar population, with 
ages of 8-12\,Gyr, that extend at least to $\rm [Fe/H]\approx0$.  
This population is not only old, but also its stars have kinematic properties 
typical of the Galactic thick disk, and chemical 
properties similar to what is found in the Galactic thick disk. 
Furthermore, preliminary results show that the
abundance and age trends do not vary with
either of the $\ulsr$, $\vlsr$, and $\wlsr$ velocities
\citep[Bensby et al.~in prep., but see also][]{bensby2006venice_short}.
Therefore, this appears to be manifest evidence that this stellar 
population indeed is the Galactic thick disk.
That the thick disk really reach all the way up solar metallicities verifies
the existence of the ``knee'' present in most
thick disk $\rm [\alpha/Fe]$ trends.
Hence the thick disk
formed stars for at least 3\,Gyr and experienced strong enrichment, from both
SN\,II and SN\,Ia, during this period, ending $\sim 8$-9\,Gyr ago.

\subsection{The relation between the thin and thick disks}

In our sample, the most metal-poor stars with thin disk kinematics have
metallicities of $\rm [Fe/H]\approx-0.7$ and ages 
around 5\,Gyr. Hence, these stars are considerably younger than the 
most metal-rich thick disk stars at $\rm [Fe/H]\approx0$ whose ages 
are 8-9\,Gyr.
At super-solar metallicities, the thin disk stars appear to have ages 
comparable to those of the most metal-poor ones, i.e. $\sim5$\,Gyr, 
suggesting that they formed at the same time! This phenomenon could be 
explained by the infall of gas into the Galaxy, which initially was
poorly mixed with the remains of the old metal-rich gas. 
The first stars of the thin disk could then be 
metal-rich ($\gtrsim 0.3$), metal-poor ($\lesssim -0.5$), or, depending
on the degree of mixing of the gas, of any metallicity in the range
$\rm -0.7\lesssim [Fe/H] \lesssim +0.4$. 
This scenario may explain why there is no well-defined
age-metallicity relation in the solar neighbourhood
\citep[e.g.,][]{edvardsson1993short,feltzing2001,haywood2006}. However, we caution that age
uncertainties can be large and that the increase in dispersion of the metallicity
with stellar age, for nearby stars, partly could be due to migration of stellar 
orbits \citep[e.g.,][]{haywood2006,wielen1996}. 

Figure~\ref{fig:agefeh}b  shows [Ba/Fe] versus stellar age.
The two disks appear to follow smoothly in
time and there also appears to be a quiescent period of 1-2\,Gyr 
when almost no stars were formed, some 6-7\,Gyr ago.
However, our thin disk stellar
sample has by no means been selected to probe its oldest parts.
Hence, a possible hiatus, and the fact that that there are (a few) 
stars that have ages in betwen the two disks, should be investigated with
a sample targeted for the oldest thin disk.

Figure~\ref{fig:agefeh}b also helps to further understand the origin and
evolution of Ba in the Galactic disks. The "bump'' in the
thin disk [Ba/Fe]-[Fe/H] trend (Fig.~\ref{fig:haltplottar0}e) is no longer 
seen. As the most metal-rich thin disk stars evidently are not the youngest ones 
there is now instead a steady increase in [Ba/Fe] toward younger ages. 
The first, flat portion of the [Ba/Fe] trend is consistent with being 
due to the $r$-process. As the $s$-process becomes
significant (due to AGB stars), [Ba/Fe] will rise. 
The position of the Sun is consistent with an origin during the
early times of the thin disk, when Ba enrichment 
was mainly $r$-process dominated but started to give way to being 
$s$-process dominated. However, we caution that the solar Ba composition 
is thought to be $\sim80$\,\% $s$-process and $\sim20$\,\% $r$-process
\citep[e.g.,][]{arlandini1999short}.

\section{Summary}

We have presented clear evidence that 
the Galactic thick disk reaches at least solar metallicities, and thus that it 
experienced strong chemical enrichment during an early period ending some
8-9\,Gyr ago. 

The plot of [Ba/Fe] versus time, instead of [Fe/H], offers a more straightforward
interpretation of  the evolution of Ba at high metallicities.

We find that even the most metal-rich stars of the thick disk are older than
the thin disk population, with a possible hiatus in the star formation between
these two populations. We are continuing to investigate these relationships
with a stellar sample designed to target the oldest stars of the thin disk.

\acknowledgments

This work was supported by the National Science Foundation, grant AST-0448900.
SF is supported by a grant from the Knut and Alice Wallenberg Foundation.
We thank the anonymous referee for valuable comments.

\bibliographystyle{apj}
\bibliography{referenser}

\end{document}